\documentclass{article}%referee

\usepackage[utf8]{inputenc}
\usepackage[USenglish]{babel}

\usepackage{amsfonts,amssymb,amsmath,  bm}%amsthm,
\usepackage{amsthm}
\usepackage{hyperref}
\usepackage{graphics}
\usepackage{subfig}
\usepackage{xcolor}
\usepackage{multirow}
\usepackage{standalone}

\newcounter{numero} \numberwithin{numero}{section}
\theoremstyle{plain}
\newtheorem{theorem}[numero]{Theorem}

\newtheorem{property}[numero]{Property}
\newtheorem{lemma}[numero]{Lemma}

\theoremstyle{definition}
\newtheorem{definition}[numero]{Definition}
\theoremstyle{remark}

\newcommand{\Z}{\mathbb Z}

\newcommand{\R}{\mathbb R}

\newcommand{\lii}{[\![}
\newcommand{\rii}{]\!]}

\renewcommand{\d}{\mathrm d}% d droit pour les intégrales ou distance
\def\B2{\text{B}_{\|.\|_{2}}}%boule norme 2

\newcommand{\C}{\mathcal C}% courbe C
\newcommand{\A}{\mathcal A}% arc A
\newcommand{\lgth}{\mathcal L}% length
\newcommand{\ltb}{LTB}% locally turn-bounded
\newcommand{\sa}[1][a,b]{\bm{C}_{#1}}

\usepackage[normalem]{ulem} %pour barrer du texte : \sout
\newcommand{\ssout}[1]{{\scriptsize{\sout {#1}}}} % Small sout
{}
\newcommand{\eelq}{\color[rgb]{0,0,0}}%\end{elq}}
{}
\newcommand{\beb}{\color[rgb]{0,0,0}}%\begin{eb}}
\newcommand{\eeb}{\color[rgb]{0,0,0}}%\end{eb}}

\usepackage{tikz, tkz-euclide}
\usetikzlibrary{calc, intersections, through, backgrounds, topaths, shapes.misc, shapes.geometric, shapes.multipart, hobby, arrows, arrows.meta, chains}
\usetikzlibrary[angles, quotes]

\newcommand{\mytkzReportLong}[5][0]{
  \coordinate (sommet1) at ($ (#2)!#4! 90+#1:(#3) $);
  \coordinate (sommet2) at ($ (#3)!#4!-90+#1:(#2) $);
  \draw [very thin, gray] (#2)--(sommet1) (#3)--(sommet2);
  \draw [very thin, gray, Stealth-Stealth] (sommet1) to coordinate[pos=0.5, label=#5] (etiquette) (sommet2);
}

\pgfdeclareplotmark{open}{$\subset$}

\newcommand{\mytkzBezier}[2]{($#1 - #2$) .. #1 .. controls ($#1 +#2$) }

\newtheorem{theo:curvature_lipschtiz_reach}{Theorem~\ref{theo:curvature_lipschtiz_reach}}
\newtheorem{theo:increasing_subsequence_midpoints}{Theorem~\ref{theo:increasing_subsequence_midpoints}}
\newtheorem{theo:convergence_estimator}{Theorem~\ref{theo:convergence_estimator}}
\newtheorem{theo:convergence_estimator_regular}{Theorem~\ref{theo:convergence_estimator_regular}}

\usepackage{lineno,hyperref}
\modulolinenumbers[5]

\begin{document}

%\begin{frontmatter}

\title{LTB curves with Lipschitz turn are par-regular.}

% \titlerunning{Monotonic sampling of a Jordan curve from its digitization and application}

\author{\'E. Le Quentrec \and L. Mazo \and \'E. Baudrier \and M. Tajine \\
ICube-UMR 7357, 300 Bd S\'ebastien Brant - \\  
CS 10413 - 67412 Illkirch Cedex FRANCE \\
etiennelequentrec@free.fr
}

% \authorrunning{\'E. Le Quentrec et al.}

% \beb{}
% \institute{ICube-UMR 7357, 300 Bd S\'ebastien Brant -
%   CS 10413 - 67412 Illkirch Cedex FRANCE\\
%   \email{elequentrec@unistra.fr}%,\\
  % WWW home page:
  % \texttt{http://icube-miv.unistra.fr/en/index.php/???}
% }
\maketitle

Preserving the topology during a digitization process is a requirement of first importance. To this end, it is classical in Digital Geometry to assume the shape borders to be \emph{par-regular}. Par-regularity was proved to be equivalent to having positive \emph{reach} or to belong to the class $C^{1,1}$ of curves with Lipschitz derivative.
Recently, we proposed to use a larger class that encompasses polygons with obtuse angles, the \emph{locally turn-bounded} curves.
The aim of this technical report is to define the class of par-regular curves inside the class of  locally turn-bounded curves using only the notion of \emph{turn}, that is of integral curvature.
To be more precise, in a previous article,
 we have already proved that par-regular curves are locally turn-bounded. Incidentally this proof lead us to show that the turn of par-regular curves is a Lipschitz function of their length. We call the class of curves verifying this latter property the curves with \textit{Lipschitz turn}. In this technical report, we prove the converse assertion : locally turn-bounded curves with Lipschitz turn are par-regular.
The equivalence is stated in Theorem \ref{theo:curvature_lipschtiz_reach} and the converse assertion is proved in Lemma \ref{cor:curvature_lipschtiz_reach}.
In section \ref{sec:par_regularity}, we recall the definition of par-regularity and equivalently of sets with positive reach. In section \ref{sec:ltb_curves},  we present the notions of curves locally turn-bounded and of curves with Lipschitz turn.
Throughout this latter section, some of intermediate steps (Lemmas \ref{lem:turn_contains_convex}  and \ref{lem:LTBextremityDistance}) are proved just after the introduction of their related notions.
The last section (section \ref{sec:the_equivalence}) is dedicated to the proof of the equivalence of the notions.
\eelq

% \belq Je ne sais pas quoi en faire pour l'instant \eelq
% In this section, we show how the notion of \ltb{} curve is linked to the well-known concepts of curve with positive reach~\cite{Federer} and par-regular curve~\cite{Pavlidis}, that is to the $\text C^{1,1}$ regularity class (curves with Lipschitz unit tangents).
% The aim of this subsection is to prove (Theorem \ref{theo:curvature_lipschtiz_reach}) that any \ltb{} curve with Lipschitz turn  has a positive reach
%  (blue arrow  Figure \ref{fig:equivalent_notions}).
% The following theorem links \ltb{} curves with Lipschitz turn and curves with positive reach (and so, with par-regular curves).

% \begin{figure}
% \begin{center}
% \input{Fig/equivalent_notions.tex}
%  \end{center}
%  \caption{ 
%  Each rectangle corresponds to a notion. An arrow  from a notion to another one means that if a curve has the first property then it has the second one.  Notice that we lost quantitative information when using the implication from curve with positive reach to  $C^{1,1}$ curve. }
%  \label{fig:equivalent_notions}
%  \end{figure}

\section{Par-regularity and equivalent notions}
\label{sec:par_regularity}
Let us first recall the definition of reach and of par-regularity.

\begin{definition}[reach]
\label{def:reach}
 The \emph{medial axis} of a compact set $K$ is the set of points having at least two nearest neighbours in $K$. The \emph{reach} is the minimal
  distance between $K$ and its medial axis.
\end{definition}

Having positive reach is equivalent to be of class $\C^{1,1}$ (the class of curves parameterized by a $C^1$ function whose derivative is  Lipschitz)  \cite{Federer}.

Regarding par-regularity, we choose the same definition as in \cite{GrossLatecConr} and \cite{LachaudThiber}.

\begin{figure}[h!]
\begin{center}
      \begin{tikzpicture}[scale= 0.3]
            \node (example) at (-9,0){\tiny{$\times$}};
            \draw (example) circle (0.9);
            \coordinate (hexa) at ($(example)+(0,0.9)$); 
            \coordinate (bexa) at ($(example)-(0,0.9)$);
            \mytkzReportLong{hexa}{bexa}{-1cm}{left:$2r$}
            \draw[blue, thick, fill= blue!20] 
            (0, -7).. controls ($(0, -7)+(-3,0)$)  
            and \mytkzBezier{(-7,0)}{(0,3)}
            and \mytkzBezier{(-3,5)}{(1,0)}
            and \mytkzBezier{(-1,3)}{(0,-1)}
            and \mytkzBezier{(-3,0)}{(0,-1)}
            and \mytkzBezier{(0,-3)}{(2,0)}
            and \mytkzBezier{(3,0)}{(0,1)}
            and \mytkzBezier{(1,3)}{(0,1)}
            and \mytkzBezier{(3,5)}{(1,0)}
            and \mytkzBezier{(7,0)}{(0,-3)}
            and ($(0,-7)+ (3,0)$) ..(0, -7);
           \draw ($(0, -7)- (0,0.9)$) circle (0.9);
           \node at ($(0, -7)- (0,0.9)$) {\tiny{$\times$}};
           \draw ($(0, -7)+ (0,0.9)$) circle (0.9);
           \node at ($(0, -7)+  (0,0.9)$) {\tiny{$\times$}};
           
           \draw ($(0, -3)- (0,0.9)$) circle (0.9);
           \node at ($(0, -3)- (0,0.9)$) {\tiny{$\times$}};
           \draw ($(0, -3)+ (0,0.9)$) circle (0.9);
           \node at ($(0, -3)+  (0,0.9)$) {\tiny{$\times$}};
           
           \draw ($(3, 5)- (0,0.9)$) circle (0.9);
           \node at ($(3, 5)- (0,0.9)$) {\tiny{$\times$}};
           \draw ($(3, 5)+ (0,0.9)$) circle (0.9);
           \node at ($(3, 5)+  (0,0.9)$) {\tiny{$\times$}};
           
            \draw ($(-1, 3)- (0.9,0)$) circle (0.9);
           \node at ($(-1, 3)- (0.9,0)$) {\tiny{$\times$}};
           \draw ($(-1, 3)+ (0.9,0)$) circle (0.9);
           \node at ($(-1, 3)+  (0.9,0)$) {\tiny{$\times$}};
        \end{tikzpicture}
 \end{center}
 \caption{\label{fig:parregularite} Par-regularity demands that at each point of the boundary of the blue shape, there exists an inside osculating disk and an outside osculating disk both of radius $r$.}
\end{figure}
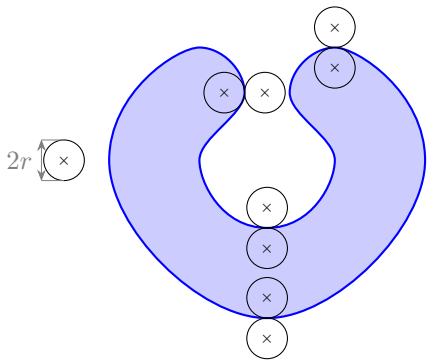

\begin{definition}[par($r$)-regularity]
\label{def:parRregularity}
Let $\C$ be a Jordan curve of interior K.
 \begin{itemize}
  \item A closed ball $\bar{B}(c_i,r)$ is an \emph{inside osculating ball} of radius $r$  at point $a \in \C$ if $\C \cap  \bar{B}(c_i,r)= \{ a\}$ and $\bar{B}(c_i, r) \subset K \cup \{a\}$.
  \item A closed ball $\bar{B}(c_e,r)$ is an \emph{outside osculating ball} of radius $r$ at point $a \in \C$ if $\C \cap  \bar{B}(c_e,r)= \{ a\}$ and $\bar{B}(c_e, r) \subset \left( \R^2 \setminus (\C\cup K) \right) \cup \{a\}
   $.
  \item A curve $\C$
   is \emph{par($r$)-regular} if there exist inside and  outside osculating balls of radius $r$ at each $a \in \C$.
  \end{itemize}
\end{definition}
The definition of par-regularity is illustrated in Figure \ref{fig:parregularite}. 

Par-regularity is equivalent to  having a  positive reach~\cite{LachaudThiber}.

\section{Locally turn-bounded curves and curves with Lipschitz turn}
\label{sec:ltb_curves}

\subsection{Turn of a curve}
\begin{figure}[h!]
    \centering
     \begin{tikzpicture}[scale= 0.5]
      \coordinate (a1) at (-2,2); \coordinate (a2) at (0,-1);
      \coordinate (a3) at (1,1); \coordinate (a4) at (3,1);
      \coordinate (a5) at (1,4); \coordinate (a6) at (0,2);
      \path (a1) to coordinate[pos=1.5] (b1) (a2) to
      coordinate[pos=1.5] (b2) (a3) to coordinate[pos=1.5] (b3) (a4)
      to coordinate[pos=1.5] (b4) (a5) to coordinate[pos=1.5] (b5)
      (a6) to coordinate[pos=1.5] (b6)(a1); \draw pic
      [draw=green!50!black, fill=green!20, angle radius=6mm] {angle =
        b6--a1--a2}; \draw pic [draw=green!50!black, fill=green!20,
      angle radius=6mm] {angle = b1--a2--a3}; \draw pic
      [draw=green!50!black, fill=green!20, angle radius=6mm] {angle =
        a4--a3--b2}; \draw pic [draw=green!50!black, fill=green!20,
      angle radius=6mm] {angle = b3--a4--a5}; \draw pic
      [draw=green!50!black, fill=green!20, angle radius=6mm] {angle =
        b4--a5--a6}; \draw pic [draw=green!50!black, fill=green!20,
      angle radius=6mm] {angle = a1--a6--b5}; \draw [dashed]
      (a2)--(b1) (a3)--(b2) (a4)--(b3) (a5)--(b4) (a6)--(b5)
      (a1)--(b6); \draw [thick] (a1) --(a2) --(a3) --(a4) --(a5)
      --(a6) --cycle;
      \draw [->] (0.8,1.5) arc[radius=3mm, start angle=0, end angle=330];
      \draw[thick, blue] plot [smooth cycle] coordinates{(a1) (a2) (a3) (a4) (a5) (a6)};
      
    \end{tikzpicture}
    \caption{The turn of the inscribed polygon is the sum of the green angles. The turn of the blue Jordan curve is the supremum of the turn of the inscribed polygons}
    \label{fig:turn_polygon}
\end{figure}
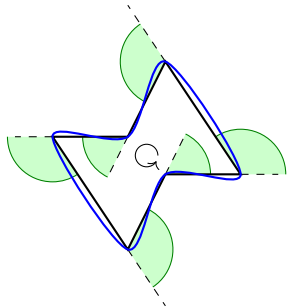

The definition of locally turn-bounded curves and curves with Lipschitz turn are both based on the notion of turn introduced by Milnor in \cite{Milnor}.
\begin{definition}[Turn of a curve {\cite{IrregularCurves}}] \hfill
\begin{itemize}
\item \emph{The turn} $\kappa(L)$ of a polygonal line $L=[x_i]_{i=0}^{N-1}$
  is defined by:
    \begin{equation*}
      \kappa(L):=\sum_{i=1}^{N-2}\angle(x_i-x_{i-1}, x_{i+1}-x_{i})\enspace.
    \end{equation*}
\item  \emph{The turn} $\kappa(P)$ of a polygon $P=[x_i]_{i \in \Z/N\Z}$
  is defined by (see Figure \ref{fig:turn_polygon} ):
    \begin{equation*}
      \kappa(P):=\sum_{i \in \Z/N \Z}{\angle(x_i-x_{i-1}, x_{i+1}-x_{i})}\enspace.
    \end{equation*}
% \item In the rest of the paper, we write $C_{a,b}$ for an arc of curve between the points $a$ and $b$; moreover, the topology on the curve and its arcs is the topology induced on the curve, therefore, an open arc $\mathring{C}$ is the arc $C$ minus its endpoints.

\item \label{defitem:chain} %
A sequence  $(a_j)$ of points of a simple closed curve $\C$ forms a \emph{chain} if
%there exists an injective parametrization $\gamma$ of $\C$ such that for each $i<j$, there exists $t_i, t_j$ such that $\gamma(t_i)=a_i$, $\gamma(t_j)=a_j$ and $t_i<t_j$.
for each pair $(i,j)$,
the intersections of the two open arcs of $\C$ from $a_i$ to $a_j$ with the set $\{a_k\}$ are exactly the subsets $\{a_k\}_{k\in\lii i+1,j-1\rii}$ and $\{a_k\}_{k\in\lii j+1, i-1\rii}$.

\item A polygonal line (or a polygon) is said to be \emph{inscribed} in $\C$ if its ordered sequence of vertices forms a chain of $\C$.
\item \label{defitem:turn} \emph{The turn} $\kappa(\C)$ of a simple curve $\C$ (respectively of a Jordan curve) is the supremum of the turn of its inscribed polygonal lines (respectively of its inscribed polygons).
\end{itemize}
\end{definition}

The properties of the turn used in rest of the report are recalled in Property \ref{prop:turn_basic}. 

\begin{property}[\cite{IrregularCurves}]
\label{prop:turn_basic}
\leavevmode
\begin{itemize}
 \item The turn coincides with the integral of the usual curvature on $C^2$ curves.
 \item  (Fenchel's Theorem) The turn of a Jordan curve is greater than or equal to $2 \pi$. The equality case occurs if and only if the interior of $\C$ is convex.
  \item Every curve of finite turn has left-hand and right-hand tangent vectors $e_l(c)$ and $e_r(c)$ at each  of its points.
  \item For any arc $C_{a,b}$ of finite turn containing a point $c$,
  $$\kappa(C_{a,b})=\kappa(C_{a,c})+ \kappa(C_{c,b}) + \angle(e_l(c), e_r(c)).$$
 \item For any Jordan curve $\C$ of finite turn containing a point $c$,
  $$\kappa(\C)=\kappa(\C \setminus \{c\})+  \angle(e_l(c), e_r(c)). $$
\end{itemize}

\end{property}

The aim of the report is to characterize par-regularity in terms of turn. 
Par-regular curves are defined by the property of bypassing disks of controlled radius. 
The following lemma will permit us to transfer the knowledge of curvature of such circles to the curvature of the par-regular curve. 
% \ssout{In Lemma \ref{cor:curvature_lipschtiz_reach}, the arc $\C'$ of Lemma \ref{lem:turn_contains_convex} is chosen to be an arc of circle.} 
The lemma is a slight improvement of Lemma 2 \cite{2-LMBT20}. This version is nevertheless necessary to get the result.
The proof remains essentially the same.
\begin{lemma}
\label{lem:turn_contains_convex} 
 Let $\C$ be a curve with endpoints $a$, $b$ such that the straight segment $(a,b)$ does not intersect the curve $\C$.
  Let $\C'$ be a simple curve  from $a$ to $b$ such that $\C'$ lies in the closure of the interior of the Jordan curve $\C\cup[a,b]$ and $\C'\cup[a,b]$ is convex.
  Then $\kappa(\C) \geq \kappa(\C')$.
\end{lemma}
% \blm{}Fixer ET informer notation half-line : parfois $[a,b)$ (ambigu), parfois $[a, b$ \elm{}
\begin{proof}
 Throughout the proof, the half-line with initial point $a$ and passing through $b$ will be noted $\overrightarrow{ \rm ab }$.
Firstly, assume that $\C'$ is a polygonal line.
We set $\C':=[a,p_1,\dotsc ,p_m,b]$.
Let $c$ be any point in $(a,b)$. 
% and $Q=[a,q_1,$ $\dotsc ,q_{m},b]$  be the polygonal line obtained by projecting from the point $c$ on the curve $\C$ \blm{} the vertices of \elm{}the polygon $\C'$. By projection of a point $x$, we mean the first intersection point $y$ between $\C$ and the half-line $D$ starting from $c$ and directed by $x-c$. This intersection exists and is well defined for $\C'$ lies in the closure of the interior of $\C\cup[a,b]$ and $\C\cap D$ is a compact set.
% Note that we do not assert that the point $q_i$ is the projection of the point $p_i$ \belq SI JUSTEMENT \eelq 
% but we claim that the polyline $\C'$ lies in the closure of the interior of the polygon $Q\cup[a,b]$ and the polyline $Q$ is inscribed in $\C$. 
For any $i \in \lii 1, m \rii$, let $q_i$ be the first intersection between the half-line $\overrightarrow{ \rm cp_i }$ and $\C$. Let $Q$ be the polygonal line $[a, q_1,\;\dotsc ,q_{m},b]$. Let us show that $Q$ is inscribed in $\C$, i.e. by definition that the sequence of its vertices is a chain.
Assume by contradiction that $(a, q_1,$ $\dotsc ,q_{m},b)$ is not a chain of $\C$. Then there exists $(i,j,k)$ such that $i<j<k$ and $(q_i, q_j, q_k)$ is not a chain of $\C$  or equivalently $(q_i, q_k, q_j)$ is a chain of $\C$ (up to consider $q_0:=a$ and $q_{m+1}:=b$). 
Observe that this assumption in particular implies that $\C'$ has more than two vertices: $\C'\neq [a,b]$.
Therefore, the interior of $\C'\cup[a,b]$ is not empty.
% C'EST ICI QU'IL FAUDRAIT RAJOUTER LE FAIT QUE CET INTERIEUR EST INCLUS DANS L'INTERIEUR DE $\C\cup[a,b]$. CE N'EST PEUT ETRE PAS NECESSAIRE.
Let $C_{q_i,q_k}$ be the closed arc of $\C$ delimited by $q_i$ and $q_k$. 
Let $T$ be the closed angular sector delimited by the half-lines $\overrightarrow{ \rm c p_i}$ and $\overrightarrow{ \rm c p_k}$ and containing the segment $[p_i, p_k]$.  
Since $T$ contains points inside and other points outside the Jordan curve $\C \cup [a,b]$, the set $T \setminus \C$ has at least two connected components.
% \blm{}\\\small On peut admettre cela ("$T$ contains points inside and other points outside"), oui même si pour etre complet, il faudrait le prouver (l'interieur de $\C'\cup[a,b]$ est inclus dans l'intérieur de $\C\cup[a,b]$).\\ \elm{}
Let $S$ be the topological closure of the connected component of $T \setminus \C$ containing $c$. Notice that \beb{}$\partial S \subset [c,q_i] \cup C_{q_i,q_k} \cup [q_k, c]$\eeb{} since $\C$ does not intersect $[c, q_i)$ nor $[c, q_k)$.
Let us show that $p_j \in S$. 
% CE QUI SUIT ME PARAIT INUTILEMENT COMPLIQUé
%The Jordan curve $\C'\cup [a,b]$ has a convex interior,  by Fenchel's Theorem \ref{prop:turn_basic} $\kappa(\C'\cup [a,b])=  2 \pi$. 
%Since $[c, p_i, p_j, p_k]$ is a chain of $\C'\cup [a,b]$, $\kappa([c, p_i, p_j, p_k])=  2 \pi$, that is $[c, p_i, p_j, p_k]$ is a convex polygon.
Since %$[c, p_i, p_j, p_k]$ 
\beb{}$(c, p_i, p_j, p_k)$\eeb{} is a chain of the convex curve $\C'\cup [a,b]$, it defines a convex polygon.
 Then the point $p_j$ belongs to the angular sector  $T$.
 % JE TOUVE QUE LE "OTHERWISE" EST PLUS DÉRANGEANT QU'EXPLICATIF
%(otherwise the segment $[c, p_j]$ would be outside the convex polygon $[c, p_i, p_j, p_k]$).
Therefore, $p_j$ belongs to $S$.
% \blm{}
Let $q'$ be a point in the intersection between the half-line $\overrightarrow{ \rm c p_j}$ and the curve arc $C_{q_i,q_k}$.
% \elm{}. 
By its definition, the point $q_j$ belongs to $[c, q']$ and since $[c,q_j)$ does not intersect $\C$, $q_j$ belongs to $S$.
% \blm{}\\\small Suppression du cas "If $q_j=b$". En effet, dans ce cas $p_j\in(a,b)$ car, par definition, $p_j$ n'est ni le premier ni le dernier point. Du coup, $\C'=[a,b]$.\\\elm{} 
%, otherwise let $\mathcal D$ be the arc of $\C$ between $q_j$ and $b$.
Let $C_{q_j,b}$ be the arc of $\C$ between $q_j$ and $b$. 
Since the curve $\C$ is simple, the arc $C_{q_j,b}$ does \beb{}not\eeb{} intersect $C_{q_i,q_k}$. 
Moreover, the arc $C_{q_j,b}$ does \beb{}not\eeb{} intersect the half-open segments $[c, q_i)$ and $[c, q_k)$ by definition of $q_i$ and $q_k$.
Then, the arc $C_{q_j,b}$ has its end $q_j$ in $S$, its other end 
%$a$ or 
$b$ outside $S$ but does \beb{}not\eeb{} intersect $\partial S$. Contradiction! Then $(a, q_1,$ $\dotsc ,q_{m},b)$ is a chain of $\C$.

Then, $\kappa(\C) \geq \kappa(Q)$ by definition of $\kappa(\C)$, $\kappa(Q\cup[b,a])\geq\kappa(\C' \cup[b,a])$ by Fenchel's Theorem (Property~\ref{prop:turn_basic}) and
\begin{align*}
 \angle(a-b, p_1-a) &\geq \angle(a-b, q_1-a)\\
 \angle(b-a, p_m-b) &\geq \angle(b-a, q_m-b)
\end{align*}
for $\C'$ is inside $Q\cup[a,b]$.
Since
\begin{equation*}
 \kappa(\C' \cup[b,a])=\kappa(C' )+\angle(a-b, p_1-a) +\angle(a-b, b-p_m)
\end{equation*}
 and
 \begin{equation*}
 \kappa(Q\cup[b,a])=\kappa(Q)+\angle(a-b, q_1-a) +\angle(a-b, b-q_m)
 \end{equation*}
 by definition of the turn of a polygon, the result holds if $\C'$ is a polygonal line.
 If $\C'$ is not a polygonal line, then, from the first part of the proof,  $\kappa(P) \leq \kappa(\C)$ for any $P$ inscribed in $\C'$.
 By definition of the turn and since the supremum is the smallest upper bound,
 \begin{equation*}
  \kappa(\C') \leq \kappa(\C)\enspace.\quad
 \end{equation*}

\end{proof}

\begin{figure}[!ht]
   \begin{center}
        \begin{tikzpicture}
%     \clip (-4,-1) rectangle (2.3,3.4);
  \coordinate (a) at (-2,0);
  \coordinate (b) at (2,0);
  \coordinate (a1) at (0,1);
  \coordinate (a2) at (-1,2);
  \coordinate (a3) at (0,3);
  \coordinate (p1) at (-0.2,0.75);
  \coordinate (p2) at (0.3,0.85);
  \coordinate (c) at (0.3,0);

  \path[draw, thick, blue, name path = arc] 
  (a).. controls +(4,-1.5)
  and \mytkzBezier{(-3,0.5)}{(0.5,0.5)}
  and \mytkzBezier{(a1)}{(0.5,0)}
%   and \mytkzBezier{(a2)}{(-2,1)}
  and \mytkzBezier{(a3)}{(3,0)}
  and (1.5,2) .. (b);
%   and (-1.25,1) .. (a1).. controls +(0.5,0) 
%   and (0,1.5) .. (a2).. controls +(-2,1) 
%   and (-1,3) .. (a3).. controls +(3,0) 
%   and (1.5,2) .. (b);
  \path[ name path = proj1] (c) -- ($(c) + 5*($(p1)-(c)$)$) ;
  \path[ name path = proj2] (c) -- ($(c) + 5*($(p2)-(c)$)$) ;

  \node[ name intersections={of=proj1 and arc, name=q1}]{};
  \node[ name intersections={of=proj2 and arc, name=q2}]{};

  \draw[thick, blue] (a) node[left, black] {$a$} -- (b) node[right, black]{$b$};
  \draw [thin]
    (a) -- (p1) node[below, black]{$p_1$}
        -- (p2) node[right, black]{$p_2$}
        -- (b);
  \draw [dashed] (c) node[below]{c} --(q1-1) node[above left]{$q_1$};
  \draw [dashed] (c)--(q2-1) node[above right]{$q_2$};
  \draw [thin, red] (a) -- (q1-1) -- (q2-1) -- (b);
  \end{tikzpicture}
   \end{center}
  \caption{\label{fig:projection}
  Blue: the curve $\C$ and the line segment $[a,b]$.
  Black: the polygonal line $\C'=[a,p_1,p_2,b]$.
  Black, dashed: the projection of $p_1$ and $p_2$ on $\C$ yields the points $q_1$ and $q_2$.
  Red: the polygonal line $Q=[a,q_1,q_2,b]$.
  }
\end{figure}
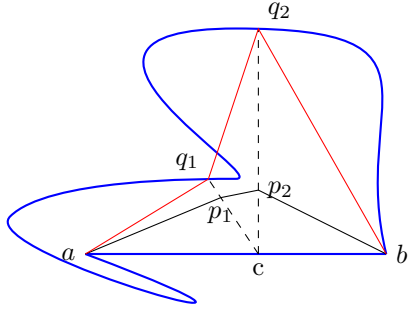

\subsection{Locally turn-bounded curves}
We introduced in~\cite{ourDGCI2019} a local geometric feature based on the turn.
It consists in locally bounding the turn of the curve in order to forbid some artifacts. This feature allows us to consider a wider class than the par-regular curves usually used for estimation in digital geometry.

\begin{definition}[LTB curves \cite{ourDGCI2019}]
\label{def:ltb}
 A Jordan curve $\C$ is \emph{$(\theta,\delta)$-locally turn-bounded} ($(\theta,\delta)$-\ltb) if,  for any two points $a$ and $b$ in $\C$ such that the Euclidean distance $\d(a,b) < \delta$, the turn of one of the arcs of the curve $\C$ delimited by $a$ and $b$  is less than or equal to $\theta$.
 A Jordan curve $\C$ is \emph{$\delta$-locally turn-bounded} ($\delta$-\ltb) if it is $(\frac\pi 2, \delta)$-LTB. 
\end{definition}

In particular, a $(\theta,\delta)$-LTB curve cannot have \emph{angular points} of turn greater than $\theta$, i.e. points $c$ for which $\angle(e_l(c), e_r(c))> \theta$ (see  \cite[Proposition 3]{ourDGCI2019}).

% \begin{figure}
% \begin{center}
% \input{Fig/two_artefacts}
% \end{center}
% \caption{The curve on the left has a thin spike avoiding all the centers of pixels.  Alike, we can build curves having arcs arbitrarily far from their Gauss digitization. Even if the curve stay close to its digitization, it can oscillate a lot around its digitization. \blm{}Such artifacts\elm{} can induce for instance
%  an arbitrarily big difference between the length of the curve and the length of its digitization.
% \label{fig:two_artefacts}
% }
% \end{figure}

\begin{property}[{\cite[Lemma 2]{2-LMBT20}}]
\label{prop:straightest_arc}
Let $\C$ be a $\delta$-\ltb{} curve. Let $a,b$ points of $\C$ such that $\d(a,b) < \delta$. Then there exists a unique arc of $\C$ delimited by the points $a$ and $b$ and whose turn is less than or equal to $\frac{\pi}{2}$. 
\end{property}

\begin{definition}[Straightest arc, {\cite[Definition 6]{2-LMBT20}}]
\label{def:straightest_arc}
Let $\C$ be a $\delta$-\ltb{} curve. Let $a,b$ points of $\C$ such that $\d(a,b) < \delta$. The unique arc of $\C$ delimited by the points $a$ and $b$ and whose turn is less than or equal to $\frac{\pi}{2}$ is called \emph{the straightest arc between $a$ and $b$} and noted $\sa$.
\end{definition}

%\begin{property}[\cite{2-LMBT20}, Definition 6 and Proposition 4]
%\label{prop:straightest_arc}
 %Let $a$, $b$ be  two distinct points of a $(\theta,\delta)$-\ltb{} Jordan curve $\C$ with $\theta\leq\pi/2$. If $\|b-a \| < \delta$, then there exists a unique arc of $\C$ between $a$ and $b$ whose turn is  less than or equal to ${\pi}/{2}$. This arc noted $\sa$ is included in the closed disk with diameter $[a,b]$ and is called \emph{the straightest arc between $a$ and $b$}.
%\end{property}

%As the $(\theta,\delta)$-LTB-curve set is growing with $\theta$, the properties established for $\theta=\theta_0$ are also available for $\theta\le\theta_0$.
 %Beside, the angle $\pi/2$ is the maximum one in $\Z^2$ for which the swollen set associated to a given digital set $D$ does not contain digital point outside of $D$ (see Fig. \ref{fig:def_\asso} for an example of swollen set).
 %NON, c'est $3\pi/4$
%The converse is false. In particular, the maximal value of the angle $\theta$ for whichthe notion of ``straightest arc'' (see below) is valid is $\pi/2$.

\begin{property}[\cite{2-LMBT20}, Proposition 5]
\label{prop:locally_connected}
Let $\C$ be a $\delta$-\ltb{} Jordan curve
% with $\theta\in(0,\pi/2]$
 and $a\in\C$.
Then, for any  $\epsilon \leq \delta$, the intersection of $\C$ with the open disk $B(a,\epsilon)$ is path-connected and is therefore an arc of $\C$.
\end{property}

From Property \ref{prop:locally_connected}, we derive
that \ltb{} curves have no  local U-turns.
\begin{property}[{\cite[Proposition 12]{2-LMBT20}}]
  \label{prop:kappa_gt_pi}
  Let  $\mathcal{C}$ be a $\delta$-\ltb{}  curve.
  Let $\gamma\colon[0, t_{M}) \to \C$ be an injective parametrization of the curve $\C$ and $t_{m}\in(0,t_M)$ be such that the arc $\gamma([0,t_{m}])$ is included in  $B(\gamma(0), \frac{\delta}{2})$.
  Then,  the restriction of the function $t \mapsto \|\gamma(t)-\gamma(0)\|$ to $[0,t_{m}]$  is increasing.
\end{property}

\subsection{Curves with Lipschitz turn}
In the framework of \ltb{} curves,  smoothness  can be expressed by a Lipschitz behavior of the turn.
\begin{definition}
 \label{def:Lipschitz_turn}
 A curve $\C$ has a $k$-Lipschitz turn if for every subarc $\A$ of $\C$,
 \begin{equation*}
  \kappa(\A) \leq k \lgth(\A).
 \end{equation*}
where $\lgth(\A)$ stands for the length of arc $\A$.  
\end{definition}
Observe that a curve $\C$ has  a $k$-Lipschitz turn if and only if the turn of any subarc of $\C$  is upper bounded by the turn of an arc of circle of radius $\frac{1}{k}$ and same length. That is why the constant $k$ will often be noted by $\frac{1}{r}$. In particular, there is no spike in a  curve with a $k$-Lipschitz turn.

One of the main idea of our characterization of par-regular curves in terms of integral curvature is to compare the ratio between the arc length and the Euclidean distance between the arc end points in the case of a curve with Lispchitz turn and in the reference case of an arc of circle.
Lemma \ref{lem:LTBextremityDistance} gives us an upper bound on the length of a curve with Lipschitz turn.
Its proof relies on the Schur's Comparison Theorem ---that we recall below--- and is similar to the one of Proposition 11 in \cite{2-LMBT20}.

\begin{property}%
%\begin{theorem}
[Schur's Comparison Theorem: \cite{DiscrDiffGeom}, p. 150]
 \label{theo:schur}
 Let $\gamma$ and $\bar{\gamma}$ be two simple curves parameterized by arc length on $[0,L]$ such that:
 \begin{itemize}
  \item $[\bar{\gamma}(0), \bar{\gamma}(L)] \cup \bar{\gamma}([0,L])$ is a convex Jordan curve,
  \item for each subinterval $I \subset [0,L]$, 
  \begin{equation*}
   \kappa(\gamma(I)) \leq \kappa(\bar{\gamma}(I)).
  \end{equation*}
 \end{itemize}
Then,
\begin{equation*}
 \|\bar{\gamma}(L)- \bar{\gamma}(0)\| \leq \|\gamma(L)- \gamma(0)\|.
\end{equation*}
%\end{theorem}
\end{property}

\begin{lemma}
\label{lem:LTBextremityDistance}
 Let $\C$ be a  $\delta$-LTB curve having a $\frac{1}{r}$-Lipschitz turn with %$\theta \in (0, \frac{\pi}{2}]$ and
 $ {\delta}\ge 2r$.
 Given two points $a$, $b$ in $\C$ such that $\|b-a\| < 2r$, the straightest arc $\sa$ from  $a$ to $b$  has its length smaller than
 $2r \arcsin\left( \frac{||b-a||}{2r} \right)$.
 \end{lemma}

 \begin{proof}
 By Property \ref{prop:locally_connected}, the intersection of the open disk $B(a, 2r)$ and $\C$ is path connected. Let $\gamma$ be the parametrization by arc length of the arc of $\C$ from $a$ to $b$  in $B(a,2r)$. Then,  $\gamma(0)=a$ and $\gamma(s_1)=b$ for some $s_1>0$.
By contradiction, assume that $s_1> s_0$ where $s_0=2r \arcsin\left( \frac{||b-a||}{2r} \right)$
 and put $c=\gamma\left(s_0\right)$.
 Let  $\bar{\gamma}$ be the parametrization by arc length of some circle of radius $r$. \\
 By hypothesis, for any subinterval $I$ of $\left[0, s_0\right]$,
 \begin{equation*}
  \kappa(\gamma(I)) \leq \frac{1}{r}|I|\enspace.
 \end{equation*}
In other words, for any subinterval $I$ of $\left[0, s_0\right]$,
 \begin{equation*}
  \kappa(\gamma(I)) \leq \kappa(\bar{\gamma}(I))\enspace.
 \end{equation*}
Hence, Schur's Comparison Theorem \cite{DiscrDiffGeom} %theo:schur
 applies:
 \begin{align*}
  \|c-a\| & \geq \left\| \bar{\gamma}\left(s_0\right) -\bar{\gamma}(0) \right\| \\
   & \geq \|b-a\|\quad\text{by definition of $s_0$ and $\bar{\gamma}$}.
 \end{align*}
 The last inequality contradicts the quasi-convexity of $s\mapsto\|\gamma(s)-\gamma(0)\|$ (Property~\ref{prop:kappa_gt_pi}).
 \end{proof}

Observe that the bound of the inequality in Lemma~\ref{lem:LTBextremityDistance}
 is sharp: the equality case holds for a circle arc.

\section{Equivalence between Lipschitz turn and par-regularity}
\label{sec:the_equivalence}
\begin{figure}[h!]
 \begin{center}
 \begin{tikzpicture}[scale= 0.5]
  \def \r {1}
  \def \delt {0.5}
  \def \ddelt {0.25}
  %arc du haut
  \coordinate (ch) at (0, \r+\ddelt);
  \coordinate (egch) at ($(ch)+(-\r,0)$);
  \coordinate (edch) at ($(ch)+(\r,0)$);
  %arc en haut à gauche
  \coordinate (cgh) at ($(egch)+( -\r,0)$);
  \coordinate (egcgh) at ($(cgh)+(-\r,0)$);
  %arc en haut à droite
  \coordinate (cdh) at ($(edch)+( \r,0)$);
  \coordinate (edcdh) at ($(cdh)+(\r,0)$);

  %arc du bas
  \coordinate (cb) at (0, -\r-\ddelt);
  \coordinate (egcb) at ($(cb)+(-\r,0)$);
  \coordinate (edcb) at ($(cb)+(\r,0)$);
  %arc en bas à gaucbe
  \coordinate (cgb) at ($(egcb)+( -\r,0)$);
  \coordinate (egcgb) at ($(cgb)+(-\r,0)$);
  %arc en bas à droite
  \coordinate (cdb) at ($(edcb)+( \r,0)$);
  \coordinate (edcdb) at ($(cdb)+(\r,0)$);
  
  \draw[very thin, gray, dashed] 
                    ($(egcgh)+(0, \r)$)--($(edcdh)+(0,\r)$)
                    (egcgh)--(edcdh)
                    ($(egcgh)+(0, -\r)$)--($(edcdh)+(0,-\r)$)
                    (egcgb) --(edcdb)
                    ($(egcgb)+(0, +\r)$)--($(edcdb)+(0,+\r)$)
                    ($-3*( \r,0)$)--($3*(\r,0)$)
                    ($(egcgb)+(0, -\r)$)--($(edcdb)+(0,-\r)$)
    ;
    \node at (ch) {\small{$\times$}};
    \node at (cgh) {\small{$\times$}};
    \node at (cdh) {\small{$\times$}};
    \node at (cb) {\small{$\times$}};
    \node at (cgb) {\small{$\times$}};
    \node at (cdb) {\small{$\times$}};
    \node at (0,0) {\tiny{$\bullet$}};
  
  \draw 
    pic[draw, thick, blue, angle radius=0.5*\r cm]{angle = egch--ch--edch}
    pic[draw, thick, blue, angle radius=0.5*\r cm]{angle = egch--cgh--egcgh}
    pic[draw, thick, blue, angle radius=0.5*\r cm]{angle = edcdh--cdh--edch}
    pic[draw, thick, blue, angle radius=0.5*\r cm]{angle = edcb--cb--egcb}
    pic[draw, thick, blue, angle radius=0.5*\r cm]{angle = egcgb--cgb--egcb}
    pic[draw, thick, blue, angle radius=0.5*\r cm]{angle = edcb--cdb--edcdb}
    ;
    \draw[thick, blue] (egcgh)--(egcgb)
                (edcdh)--(edcdb);

    \mytkzReportLong{$(egcgh)+(0, \r)$}{egcgh}{-1cm}{left:$r$}            
    \mytkzReportLong{egcgh}{$(egcgh)+(0, -\r)$}{-1cm}{left:$r$}            
    \mytkzReportLong{$(egcgh)+(0, -\r)$}{$(egcgb)+(0, \r)$}{-1cm}{left:$\delta$}
    \mytkzReportLong{$(egcgb)+(0, \r)$}{egcgb}{-1cm}{left:$r$} 
    \mytkzReportLong{egcgb}{$(egcgb)+(0, -\r)$}{-1cm}{left:$r$}
    \mytkzReportLong{$(edcdh)+(0, -\r)$}{$3*(\r,0)$}{1cm}{right:reach}
\end{tikzpicture}
\end{center}
 \caption{The blue curve ---made of half-circles of radius $r$ and straight segments---  is $\delta$-\ltb{} and has $\frac{1}{r}$-Lipschitz turn. Its reach is got at the point represented by a bullet and is equal to $\frac{\delta}{2}$ \ssout(then its radius of par-regularity is smaller than $\frac{\delta}{2}$\ssout). By choosing smaller and smaller values of $\delta$ and huge values of $r$, we get a family of curves not par($r$)-regular but whose turn is $\frac{1}{r}$-Lipschitz with arbitrarily big value of $r$ whose radius of par-regularity cannot be inferred from the Lispchitz parameter $\frac1{r}$.  }
 \label{fig:sharp_bone}
\end{figure}
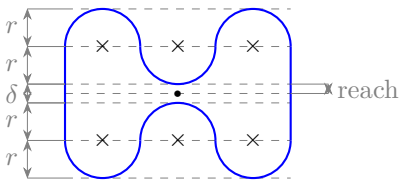

The goal of this section is to characterize par-regular curves as a subset of LTB-curves thanks to the notion of Lipschitz turn.
The example depicted in Figure \ref{fig:sharp_bone} proves that having a Lipschitz turn with known parameter $k$ is not sufficient to determine a radius of par-regularity. 
The value $\delta$ provided by the LTB-hypothesis is then useful to quantified the equivalence between the two notions.

Let us now state our equivalence theorem between Lipschitz turn and par-regularity.
\begin{theorem}
\label{theo:curvature_lipschtiz_reach} \hfill
\begin{itemize}
 \item Let $r>0$, any par($r$)-regular curve is $(\theta, 2r \sin( \frac \theta 2) )$-LTB and has a $\frac 1 r$ Lipschitz turn for $\theta \in (0, \pi]$. 
 \item Conversely, any $(\frac \pi 2, \delta)$-LTB curve with $\frac 1 r$-Lipschitz turn is par($r_1$)-regular for any $r_1 < \min (\frac \delta 2, r)$.
\end{itemize}
\end{theorem}
The first implication has already been proved in \cite[Theorem 2 and Lemma 6]{2-LMBT20}.
%\blm{}\\La def et le lemme de Schur sont utilisés en Section 4.\elm{}
% \begin{tikzpicture}
%  \def\tkze{0.2}
%  \def\tkzr{1}
%  \coordinate (g0) at ({-\tkzr-\tkze},- \tkzr);
%  \coordinate (d0) at ({\tkzr+\tkze},- \tkzr);
%  \coordinate (g1) at (-\tkze, 2*\tkzr);
%  \coordinate (d1) at (\tkze, 2*\tkzr);
% 
%  \draw[thick, blue, fill= blue!20] (g0)--(d0)
%  (g0) arc [start angle=270, end angle=90,radius=\tkzr]
%       arc [start angle=-90, end angle=90,radius=\tkzr]
%       --++(-2*\tkzr, 0)
%       arc [start angle=90, end angle=180,radius=\tkzr]
%       --++(0, -4*\tkzr)
%       arc [start angle=180, end angle=270,radius=\tkzr]
%       --++(6*\tkzr+ 2*\tkze, 0)
%       arc [start angle=-90, end angle=0,radius=\tkzr]
%       --++(0, 4*\tkzr)
%       arc [start angle=0, end angle=90,radius=\tkzr]
%     --++(-2*\tkzr, 0)
%       arc [start angle=90, end angle=270,radius=\tkzr]
%     arc [start angle=90, end angle=-90,radius=\tkzr]
%  ;
%     \mytkzReportLong{}
%     \node at (g1) {\tiny{$\bullet$}};
%     \node at (d1) {\tiny{$\bullet$}};
% \end{tikzpicture}
% The proof of Theorem~\ref{theo:curvature_lipschtiz_reach} needs \beb{} 3 lemmae\eeb{}. 
% The first lemma bounds the length of a straightest arc of a \ltb{} curve having a $\frac{1}{k}$ Lipschitz turn.
%============================
 %The proof of the ``if'' part of Theorem~\ref{theo:curvature_lipschtiz_reach} still needs 
%
%========================
From Lemmas~\ref{lem:LTBextremityDistance} and~\ref{lem:turn_contains_convex}, \beb{} we derive in Lemma~\ref{cor:curvature_lipschtiz_reach} the converse implication of Theorem \ref{theo:curvature_lipschtiz_reach}.
\begin{lemma}
\label{cor:curvature_lipschtiz_reach}
 Let $\C$ be a $\delta$-LTB curve having a $\frac{1}{r}$-Lipschitz turn with $\delta>0$. %$\theta \in (0, \frac{\pi}{2}]$.
 Then the reach of $\C$ is greater than or equal to $\min( \frac{\delta}{2},r)$.
\end{lemma}

\begin{proof}
 By contradiction assume that $\operatorname{reach}(\C)<r_1=\min(\frac{\delta}{2},r)$.
 Then there exist a point $o$ on the medial axis of $\C$ and two points $a$ and $b$ on $\C$ such that $d(o,a) = \d(o,b)=\d(o,\C) < r_1$.
 Thus, $\| a-b\| < 2r_1$.
  Let $\sa$ be the straightest arc of $\C$ between $a$ and $b$.
  On the   one hand, by Lemma \ref{lem:LTBextremityDistance} (noting that  $\C$ is $(1/r_1)$-Lipschitz for $r_1\le r$),
 \begin{equation*}
  \lgth(\sa) \leq 2r_1\arcsin\left(\frac{\|b-a\|}{2r_1}\right).
 \end{equation*}
In other words, since the sine function is increasing on $[0,\pi/2]$,
 \begin{equation}
 \label{eq:AB_greater}
 \|b-a\| \geq 2r_1 \sin \left( \frac{\lgth(\sa)}{2r_1}\right).
 \end{equation}
 On the other hand,
let  $r^{\prime}=\d(o,a)$ ($r^{\prime}<r$).
By definition of $o$, $a$ and $b$, the curve $\C$ does not intersect the interior of the circle of center $o$ and radius $r^{\prime}$.
So, let $\bar{C}$ be the arc of this last circle bounded by $\sa \cup [oa] \cup [ob]$.
%  the arc $\bar{\C}$ lies in  the compact set bounded by $\C$.
 Since  $\bar{C}\cup[a,b]$ is convex, by Lemma \ref{lem:turn_contains_convex}  we have,
 \begin{equation*}
  \kappa(\bar C) \leq \kappa(\sa).
 \end{equation*}
 Hence,
\begin{equation}
\begin{split}
\|b-a\|&= 2 r^{\prime} \sin(\frac{\kappa(\bar C)}{2}) \\
  & \leq 2 r^{\prime} \sin(\frac{\kappa( \sa)}{2}).
  \label{eq:AB_smaller}
  \end{split}
\end{equation}
The last inequality holds for the sine function is increasing on $[0,\pi/2]$ and $\kappa(\sa)\leq\pi$ because $\sa$ is the straightest arc between $a$ and $b$.

By Inequality \ref{eq:AB_greater} and Inequality \ref{eq:AB_smaller},
\begin{equation*}
 2r_1 \sin \left( \frac{\lgth(\sa)}{2r_1}\right) \leq 2 r^{\prime} \sin(\frac{\kappa( \sa)}{2}) < 2 r_1 \sin(\frac{\kappa( \sa)}{2}).
\end{equation*}
Thus,
\begin{equation*}
\frac1{r}\lgth(\sa) \le \frac1{r_1}\lgth(\sa) < \kappa(\sa),
\end{equation*}
which contradicts the Lipschitz hypothesis.\hfill 
\end{proof}

The bound  of \beb{} Lemma\eeb{}~\ref{cor:curvature_lipschtiz_reach} is sharp. Indeed, if $\C$ is a circle of radius $r$ then the reach is exactly $r$ and the reach of the $\delta$-\ltb{} curve depicted in Figure \ref{fig:sharp_bone} is exactly $\delta/2$.
% for the curve $\C$ made of half-circles and straight segments

Notice that we have proved a ``qualitative equivalence'' between the notions of positive reach and \ltb{} curve with Lipschitz turn but we failed to obtain a ``quantitative equivalence''.
Indeed, starting from a par($r$)-regular curve $\C$ and
 applying  \cite[Theorem 2 and Lemma 6]{2-LMBT20},
 we derive that, for any $0\le\theta\le\pi/2$,  $\C$ is a $(\theta, 2r\sin(\frac{\theta}{2}))$-\ltb{} curve having a $({1}/{r})$-Lipschitz turn.
 Then, from Lemma~\ref{cor:curvature_lipschtiz_reach}, we get
  that $\C$ is  a par($r\sin(\frac{\theta}{2})$)-regular curve with $\theta \leq \frac{\pi}{2}$.
  Hence, at best $\C$ is proved to be par(${\sqrt{2}}/{2}r$)-regular.
  We do not retrieve the starting parameter.

\section{Conclusion}
Theorem \ref{theo:curvature_lipschtiz_reach} permits to split the definition of par-regularity into two parts: a control of the curvature of the boundary (a par-regular curve has a Lipschitz turn) and a thickness of shape bounded by the curve (a par-regular curve is locally turn-bounded). Other generalizations of par-regularity exist in the literature. If, in \cite{LeQuentrecEtAl2021}, we have already established the link with the quasi-regularity introduced in \cite{NgoEtAl2017}, we also consider to prove that LTB curves have a positive $\mu$-reach (a generalization of reach for polygonal and regular curves) \cite{ChazalEtAl2009, ChazalEtAl2009a} .

\bibliographystyle{alpha}
\bibliography{biblio}

\end{document}